\documentclass[twocolumn,nofootinbib,prd,aps,superscriptaddress,tightenlines]{revtex4}
\usepackage{graphicx}
\usepackage{bm}
\usepackage{epsfig}

\newif\ifpdf
\ifx\pdfoutput\undefined
\pdffalse 
\else
\pdfoutput=1 
\pdftrue
\fi

\def\Slash#1{{#1\!\!\!\slash}}

\def\OMIT#1{}

\def\lqcd{\Lambda_{\rm QCD}}
\def\model{{\rm{mod}}}
\def\gev{{\rm GeV}}

\newcommand{\nn}{\nonumber} 

\newcommand{\bn}{{\bar n}}
\newcommand{\bea}{\begin{eqnarray}}
\newcommand{\eea}{\end{eqnarray}}

\newcommand{\mcdot}{\!\cdot\!}

\begin{document}
\ifpdf
\DeclareGraphicsExtensions{.pdf, .jpg}
\else
\DeclareGraphicsExtensions{.eps, .jpg}
\fi


\preprint{ \vbox{\hbox{UCSD/PTH 02-12}\hbox{UTPT 02-07}\hbox{TTP02-06} \hbox{hep-ph/0205150}  }}

\title{
Subleading shape functions in $B \to X_u \ell \bar \nu$ and the determination of $|V_{ub}|$
}

\author{Christian W.~Bauer}
\affiliation{Department of Physics, University of California at San Diego,
	La Jolla, CA 92093\footnote{Electronic address: bauer@physics.ucsd.edu}}

\author{Michael Luke}
\affiliation{Department of Physics, University of Toronto, 60 St. George St. 
	Toronto, Ontario, M5S-1A7, Canada\footnote{Electronic address: luke@physics.utoronto.ca}}

\author{Thomas Mannel}
\affiliation{Institut f\"{u}r Theoretische Teilchenphysik,
Universit\"{a}t Karlsruhe  D--76128 Karlsruhe, Germany\footnote{Electronic address: tm@particle.uni-karlsruhe.de}}

\date{\today\\ \vspace{.2cm} }

\begin{abstract}
We calculate the subleading twist contributions to the endpoint of the inclusive lepton energy spectrum in $B \to X_u \ell \bar\nu$. We show that the same two subleading twist functions that appear in the decay $B \to X_s \gamma$ govern the subleading effects in this decay. Using these results we find large ${O}(\lqcd/m_B)$ corrections to the determination of $|V_{ub}|$ from the endpoint of the charged lepton spectrum.  Using a simple model for the relevant subleading shape functions, we estimate the uncertainty in $|V_{ub}|$ from $\lqcd/m_b$ corrections  to be at the $\sim 15\%$ level for a lower lepton energy cut of 2.2 GeV.

\end{abstract}

\maketitle


A precise determination of the CKM matrix element $|V_{ub}|$ is one of the most important measurements in the current quest to overconstrain the unitarity triangle to test the standard model.   The theoretically cleanest determination of $|V_{ub}|$ comes from the  inclusive rate for $B \to X_u \ell \bar \nu$, which may be calculated model-independently via an operator product expansion in terms of local operators \cite{HQETOPE}.   Such a measurement is hampered by the experimental necessity to suppress the overwhelming charm background, which usually requires tight kinematic cuts.  With a few exceptions \cite{qsqspec}, such cuts invalidate the local OPE usually used to calculate  the inclusive branching ratio, and a twist expansion has to be performed instead \cite{twistexp}. At leading order in this twist expansion, and ignoring perturbative corrections, the differential lepton energy spectrum in inclusive $B \to X_u \ell \bar \nu$ decay is given by 
\begin{eqnarray}\label{Ellead}
\frac{d\Gamma}{dE_\ell} &=& \frac{G_F^2 |V_{ub}|^2 m_b^4}{96 \pi^3}\int \!d\omega\, \theta(m_b-2E_\ell-\omega) f(\omega)\,,
\end{eqnarray}
where $f(\omega)$ is the nonperturbative light cone structure function of the $B$ meson
\begin{eqnarray}
f(\omega) = \frac{1}{2m_B} \langle B | \bar h \, \delta(i n \cdot D + \omega) \, h | B \rangle\,.
\end{eqnarray}
Since this function is a property of the $B$ meson, it is process independent and can thus be measured elsewhere. The best way to measure this structure function is from the photon energy spectrum of the inclusive decay $B \to X_s \gamma$ \cite{shapeuniversal}. Up to perturbative and subleading twist corrections, this spectrum is directly proportional to the structure function,
\begin{eqnarray}\label{diffEgamma}
\frac{d\Gamma}{dE_\gamma} &=&  \frac{G_F^2 |V_{tb}V_{ts}^*|^2 \alpha |C_7^{\rm eff}|^2m_b^5}{32 \pi^4} f(E_\gamma)\,.
\end{eqnarray}
Thus, combining data on $B \to X_s \gamma$ with data from $B \to X_u \ell \bar \nu$, one can eliminate the dependence on the structure function and therefore determine $|V_{ub}|$ with no model dependence at leading order \cite{twistexp,shapeuniversal,ira}. The relation is
\begin{eqnarray}\label{treeVub}
\left|\frac{V_{ub}}{V_{tb}V_{ts}^*}\right|^2 = \frac{3 \alpha}{\pi} |C_7^{\rm eff}|^2 \frac{\Gamma_u(E_c)}{\Gamma_s(E_c)} + {O}(\alpha_s) + {O}\!\left(\frac{\lqcd}{m_B}\right)
\end{eqnarray}
where 
\begin{eqnarray}\label{GuGsdefs}
\Gamma_u(E_c) &\equiv& \int_{E_c}^{m_B/2} d E_\ell \frac{d \Gamma_u}{d E_\ell}\nn\\
\Gamma_s(E_c) &\equiv& \frac{2}{m_b} \int_{E_c}^{m_B/2} d E_\gamma (E_\gamma - E_c)\frac{d \Gamma_s}{d E_\gamma} \,.
\end{eqnarray}
Recently, the CLEO collaboration \cite{CLEObtou} measured the rate for $B\to X_u\ell\bar\nu$ decay in the endpoint interval $2.2\,\gev<E_\ell<2.6\,\gev$ and combined this with their measurement of the $B\to X_s\gamma$ photon spectrum to determine the value
\begin{equation}
|V_{ub}|=(4.08\pm 0.34\pm0.44\pm 0.16\pm 0.24)\times 10^{-3}
\end{equation}
where the first two uncertainties are experimental, and the third corresponds to the theoretical uncertainty in the relation between $\Gamma(B\to X_u\ell\bar\nu)$ and $|V_{ub}|$.  The fourth uncertainty is an estimate of the nonperturbative uncertainties in the relation (\ref{treeVub}), and is determined by varying the parameters of $f(\omega)$ by 10\% (as expected for an $O(\lqcd/m_b)$ correction).

In this paper we calculate the subleading twist corrections to Eq.~(\ref{treeVub}).
While perturbative corrections to this relation have been studied in detail \cite{shapeuniversal,ira}, much less is known about nonperturbative corrections suppressed by powers of $\lqcd/m_b$. In \cite{subleading} we introduced a formalism (similar to that used for higher twist calculations in deep inelastic scattering \cite{ellis}) which allows these subleading twist corrections to be parameterized in terms of subleading shape functions, and calculated the subleading twist corrections to the photon energy spectrum in $B \to X_s \gamma$. These corrections are suppressed by one power of $\lqcd/m_b$ and can be written in terms of three structure functions
\begin{eqnarray}\label{subleadingbsg}
\frac{m_b}{\Gamma_0^s}\frac{d\Gamma}{dE_\gamma} &=& (4 E_\gamma-m_b)F(m_b-2E_\gamma)
\nn\\
&&
 + \frac{1}{m_b} \left[h_1(m_b-2E_\gamma) + H_2(m_b-2E_\gamma) \right]
\end{eqnarray}
where 
\begin{eqnarray}\label{Gamma0s}
\Gamma_0^s = \frac{G_F^2 |V_{tb}V_{ts}^*|^2 \alpha |C_7^{\rm eff}|^2 m_b^5}{32 \pi^4}
\end{eqnarray}
and the functions $F$, $h_1$ and $H_2$ will be defined later. 

\begin{figure}
\includegraphics[width=8cm]{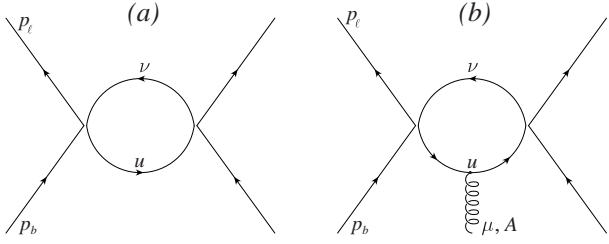}
\caption{The two full theory graphs required for the matching onto the subleading shape functions \label{fig1}}
\end{figure}

The subleading twist corrections to the lepton energy spectrum in $B \to X_u \ell \bar \nu$ may be calculated in much the same way as the photon spectrum for $B\to X_s\gamma$.
The matching calculation is performed by expanding full QCD matrix elements to subleading twist and identifying the result with non-local operators. The operator mediating the decay $B \to X_u \ell \bar\nu$ is given by
\begin{eqnarray}\label{operator}
O = (\bar u b)_{V-A} (\bar\ell \nu)_{V-A} = (\bar u \nu)_{V-A} (\bar\ell b)_{V-A}\,,
\end{eqnarray}
where we have used a Fierz transformation to obtain the right hand side of Eq.~(\ref{operator}). Using the optical theorem, the differential lepton energy spectrum can be obtained from the imaginary part of the forward scattering matrix element
\begin{eqnarray}
\frac{d \Gamma}{d E_\ell} \sim \int \!d[P.S.]\, \int \!d^4\, x e^{i q \cdot x}
\, {\rm Im} \langle B| T\{O^\dagger(x) O(0)| B \rangle .\,\,
\end{eqnarray}
The phase space integrals are automatically performed by taking the imaginary part of one loop diagram shown in  in Fig.~\ref{fig1}(a). 
Calculating this diagram and taking the imaginary part we find
\begin{eqnarray}\label{Azero}
A_a = \frac{16 \Gamma_0 E_\ell}{m_b^4}\theta(p^2) ( p^\mu p^\nu - p^2 g^{\mu\nu}) L_{\mu\nu}\,,
\end{eqnarray}
where 
\begin{eqnarray}\label{Gamma0u}
\Gamma_0 = \frac{G_F^2 |V_{ub}|^2 m_b^5}{192 \pi^3}
\end{eqnarray}
and we further defined $p^\mu = p_b^\mu - p_\ell^\mu$. Finally, $L_{\mu\nu}$ contains the contraction of the lepton spinors
\begin{eqnarray}
L_{\mu\nu} = \gamma_\mu P_L \,\Slash{p}_\ell \,\gamma_\nu P_L\,,
\end{eqnarray}
with $P_L = (1-\gamma^5)/2$. 

To expand this expression to subleading twist, we identify the scaling of the various components of the momenta. Defining 
\begin{eqnarray}
p_b^\mu = m_b v^\mu + k^\mu\,, \quad  Q^\mu = m_b v^\mu - p_\ell^\mu\,,
\end{eqnarray}
as well as the light cone vectors $n^\mu$,  $\bn^\mu$ ($n^2=\bn^2 = 0$, $n\cdot \bn = 2$, $2v = n + \bn$) by
\begin{eqnarray}
p_\ell^\mu = E_\ell \bn^\mu\,,
\end{eqnarray}
we find
\begin{eqnarray}
\bn \cdot Q &=& m_b\\
n \cdot Q &=& m_b - 2 E_\ell \sim O(\lqcd)\nn\\
k^\mu &\sim& O(\lqcd)\nn
\end{eqnarray}
(this should be compared with the power counting for the local OPE, in which $n\cdot Q$ scales as $m_b$).
Using this momentum scaling in Eq.~(\ref{Azero}) and expanding in powers of $\lqcd/m_b$ we find at leading order
\begin{eqnarray}
A_a^{(0)} = 4 \Gamma_0 \Slash{n} P_L \theta(m_b - 2 E_\ell + n \cdot k)\,.
\end{eqnarray}
For pseudoscalar $B$ meson decays, only the parity even part of this amplitude contributes. 
The matching onto the low energy theory is obtained by comparing this with the general expression at leading twist
\begin{eqnarray}
\int \! d \omega \, C_0(\omega) O_0(\omega) &=& \frac{1}{2}{\rm Tr} 
\left\{ P_+ A_a^{(0)}\right\}\nn\\
&=& 2 \Gamma_0 \, \theta(m_b - 2 E_\ell + n \cdot k)\,,
\end{eqnarray}
where $P_+ = (1 + \Slash{v})/2$ and 
\begin{eqnarray}
O_0(\omega) &=& \bar h \delta(i n \cdot D + \omega) h\,.
\end{eqnarray}
Thus, we find
\begin{eqnarray}
C_0(\omega) = 2 \, \Gamma_0 \, \theta(m_b - 2 E_\ell - \omega)\,,
\end{eqnarray}
in agreement with Eq.~(\ref{Ellead}). 

At subleading twist there are five nonlocal operators contributing to $B$ meson decays \cite{subleading}
\begin{widetext}
\begin{eqnarray}
O_1^\mu(\omega)&=& 
\bar h_v \left\{i D^\mu,\delta(in \mcdot D + \omega)\right\} h_v
\nn\\
P_2^{\mu\alpha}(\omega) &=&
\bar h_v s^\alpha \left[i D^\mu,\delta(in \mcdot D + \omega)\right] h_v \nn\\
O_T (\omega) &=& i \int\! d^4y \, \frac{1}{2 \pi} \int\! dt \,
  e^{-i\omega t} T \left( \bar{h}_v (0) h_v (t) O_{1/m} (y)
  \right)\\\
O_3(\omega) &=&
\!\!\int \!d \omega_1 d \omega_2 \, \frac{\delta(\omega-\omega_1)- \delta(\omega-\omega_2)}{\omega_1 - \omega_2} 
\Big[ h_v \delta(in \mcdot D \!+\! \omega_2) (i D_\perp)^2 \delta(in \mcdot D \!+\! \omega_1) h_v\Big]\nn\\
P_4^\alpha (\omega) &=& g
\!\!\int \!d \omega_1d \omega_2\,\frac{\delta(\omega-\omega_1)- \delta(\omega-\omega_2)}{\omega_1 - \omega_2} 
\Big[ h_v s^\alpha \delta(in \mcdot D \!+\! \omega_2) \epsilon_\perp^{\mu\nu} G^\perp_{\mu\nu} \delta(in \mcdot D \!+\! \omega_1) h_v\Big]\,,\nn
\end{eqnarray}
where $s^\alpha = P_+ \gamma^\alpha \gamma^5 P_+$. The Feynman rules for these operators were given in $n \cdot A = 0$ gauge in Ref.~\cite{subleading} (note that we have used slightly different definitions of $O_3$ and $P_4$ from Ref.~\cite{subleading}).
To match onto these operators we expand the full theory calculations to one order higher in $k^\mu/m_b$ and $n\cdot Q/m_b$. For the graph with no external gluon we find from (\ref{Azero}) \begin{eqnarray}\label{zerogluon_match}
A_a^{(1)} &=&
4 \Gamma_0 \left\{
\frac{k_\perp^2}{m_b} \Slash{n} P_L \delta (n \cdot k +m_b-2 E_\ell)
+ \left[ 2\, \frac{\Slash{k}_\perp}{m_b} P_L 
- 2 \,\frac{n \cdot k+m_b-2 E_\ell}{m_b} (\Slash{n}-\Slash{\bar{n}}) P_L \right] \theta(n \cdot k +m_b-2 E_\ell)\right\}
\end{eqnarray}
\end{widetext}
where we have defined 
\begin{eqnarray}
k_\perp^\mu = k^\mu - n^\mu\frac{\bn \mcdot k}{2} - \bn^\mu\frac{n \mcdot k}{2} \,.
\end{eqnarray} 
Of the above operators only $O_1^\mu(\omega)$ and $O_3(\omega)$ contribute to the zero gluon matrix element. In addition, we use the spinor expansion relating the full QCD spinors $u$ to the two component HQET spinors $h$
\begin{eqnarray}
u = \left(1 + \frac{\Slash{k}}{2m_b} \right) h\,,
\end{eqnarray}
and as before only the parity even part of this amplitude contributes to the decay rate. This allows us to write
\begin{eqnarray}\label{zerogluon_match1}
&&\!\!\!\!\!
\int \! d \omega \, \left[C_{1\mu}(\omega) O_1^\mu(\omega) + C_{3}(\omega) O_3(\omega) \right]\\
&&\hspace{1cm}
= 2 \frac{\Gamma_0}{m_b} \left[n \mcdot k \, \theta(n \mcdot k + m_b-2 E_\ell) 
\right.\nn\\
&&\hspace{2.5cm}\left.
+k_\perp^2 \delta(n \mcdot k + m_b-2 E_\ell)\right]\,.\nn
\end{eqnarray}
From this we can easily read off the Wilson coefficients for the subleading twist operators $O_1^\mu(\omega)$ and $O_3(\omega)$. We find
\begin{eqnarray}
C_1^\mu(\omega)  &=& \frac{\Gamma_0}{m_b}n^\mu \theta(m_b-2 E_\ell-\omega) \\
C_3(\omega) &=& -\frac{2 \Gamma_0}{m_b} \theta(m_b-2 E_\ell-\omega)\,.
\end{eqnarray}
The result for the the coefficient $C_3(\omega)$ agrees with the prediction from reparameterization invariance, which relates this coefficient to the leading order Wilson coefficient to all orders in perturbation theory \cite{mannel_RPI}
\begin{eqnarray}\label{RPI}
C_3(\omega) = - C_0(\omega)\,.
\end{eqnarray}

To determine the matching coefficients of the operators $P_2^{\mu\alpha}(\omega)$ and $P_4^\alpha(\omega)$, we need to calculate the matrix element with an explicit gluon in the final state. The required diagram is shown in Fig.~\ref{fig1}(b). Performing the calculation in $n \cdot A = 0$ gauge and expanding to subleading twist gives
\begin{widetext}
\begin{eqnarray}\label{onegluon_match}
A_1 \!\!&=&\!\! 4 \Gamma_0 \left\{
\frac{1}{m_b} \Slash{\epsilon}_\perp P_L \left[ \theta(n \mcdot k +m_b-2 E_\ell)  + \theta(n \mcdot k + n \cdot l+m_b-2 E_\ell) \right]
- \left[
\frac{1}{m_b} \Slash{n} P_L (2 k+l)_\perp\mcdot \epsilon_\perp 
\right.\right.\\
&&\left.\left.
+ \frac{2 n \mcdot k + n \mcdot l +2m_b-4 E_\ell}{m_b} \Slash{\epsilon}_\perp \!P_L
- \frac{3}{m_b} \Slash{n} P_L i \epsilon_\perp^{\mu\nu} l^\perp_\mu \epsilon^\perp_\nu \right]
\frac{\theta(n \mcdot k +m_b-2 E_\ell)-\theta(n \mcdot k + n \mcdot l+m_b-2 E_\ell)}{n \cdot l}\right\}\,,\nn
\end{eqnarray}
\end{widetext}
where
\begin{eqnarray}
\epsilon_\perp^{\mu\nu} = \epsilon^{\mu\nu\alpha\beta} v_\alpha n_\beta\,.
\end{eqnarray}
The first and second term in this expression correctly reproduce the one gluon terms from the operators $O_1^\mu (\omega)$ and $O_3(\omega)$. The Wilson coefficients for the operators $P_2^{\mu\alpha}(\omega)$ and $P_4^\alpha(\omega)$ are determined from this expression to be
\begin{eqnarray}
D_2^{\mu,\alpha}(\omega)  &=& -\frac{\Gamma_0}{m_b} i \epsilon_\perp^{\mu\alpha}  \theta(m_b-2E_\ell-\omega)\\
D_4^\alpha(\omega) &=& \frac{3\Gamma_0}{m_b} n^\alpha \theta(m_b-2E_\ell-\omega)\,.
\end{eqnarray}

To calculate the differential decay rate to subleading twist, we need expressions for the matrix elements of these operators. These matrix elements may be parameterized in terms of four subleading structure functions \cite{subleading}
\begin{eqnarray}
\langle H | O_0( \omega) | H \rangle &=& 2m_B \, f(\omega) \nn\\
\langle H | O_1^\mu( \omega) | H \rangle &=& -4m_B\, \omega f(\omega)\left( v^\mu - n^\mu \right) \nn\\
\langle H | O_3( \omega) | H \rangle &=& 2m_B \, G_2(\omega)\\
\langle H | O_T( \omega) | H \rangle &=& 2m_B \, t(\omega) \nn\\
\langle H | P_2^{\alpha,\mu}( \omega) | H \rangle &=& 
- 2m_B \, i \epsilon_\perp^{\mu\alpha}h_1(\omega)  \nn\\
\langle H | P_4^\alpha( \omega) | H \rangle &=& 4m_B \, (v^\alpha - n^\alpha)H_2(\omega)\nn
\end{eqnarray}
There are some additional simplifications that may be used to reduce the number of unknown functions. First, the function $t(\omega)$ comes from expanding the Lagrangian to higher order in $1/m_b$ and therefore always occurs in the same linear combination $f(\omega) + t(\omega)/2m_b$. 
Together with the reparametrization invariance constraint (\ref{RPI}), it therefore follows that for any $B$ meson decay one always finds the same linear combination of leading and subleading shape functions
\begin{eqnarray}
F(\omega) = f(\omega) + \frac{1}{2m_b} \left[ t(\omega) - 2  \, G_2(\omega) \right]\,.
\end{eqnarray}
Putting all this information together, we therefore find for the lepton energy spectrum in $B \to X_u \ell \bar\nu$ 
\begin{eqnarray}\label{finalbtou}
\frac{d \Gamma}{d E_\ell} &=&
\frac{2\,\Gamma_0}{m_b} \int \!d \omega \, \theta(m_b-2E_\ell-\omega) \, 
\left[ F(\omega) \left(1-\frac{\omega}{m_b} \right)
\right.\nn\\
&&\left.
- \frac{1}{m_b} h_1(\omega) 
+ \frac{3}{m_b} H_2(\omega) \right] + O\left(\frac{\lqcd^2}{m_b^2}\right)\,.
\end{eqnarray}
Eq.~(\ref{finalbtou}) is the central result of this paper. We can check this result against the known result from the local OPE by using the expansion of the nonperturbative functions \cite{subleading}
\begin{eqnarray}\label{nonlocalexpand}
f(\omega)
&=& \delta(\omega) - \frac{\lambda_1}{6} \delta''(\omega) - \frac{\rho_1}{18} \delta'''(\omega) + \ldots\nonumber\\
G_2(\omega)
&=& - \frac{2 \lambda_1}{3} \delta'(\omega) + \ldots \nonumber\\
t(\omega) 
&=& - \left(\lambda_1+ 3 \lambda_2\right) \delta'(\omega) + \frac{\tau}{2
} \delta''(\omega) + \ldots\nonumber\\
h_1(\omega) 
&=& \lambda_2 \, \delta'(\omega) + \frac{\rho_2}{2} \delta''(\omega) + \ldots\nonumber\\
H_2(\omega)
&=& - \lambda_2 \, \delta'(\omega) + \ldots \,,
\end{eqnarray}
where $\ldots$ denote terms ${\cal O}(\lqcd^2/m_b^2)$ in the twist expansion. 
This gives
\begin{eqnarray}
\frac{d \Gamma}{dy} &=& \Gamma_0 \bigg[ 2 \theta(1-y) - \frac{\lambda_1}{3m_b^2} \delta'(1-y) - \frac{\rho_1}{9m_b^3} \delta''(1-y)
\nn\\\nn\\
&&
- \frac{\lambda_1}{3m_b^2} \delta(1-y) - \frac{11\, \lambda_2}{m_b^2} \delta(1-y) 
- \frac{\rho_1}{3m_b^3} \delta'(1-y) 
\nn\\
&&
- \frac{\rho_2}{m_b^3} \delta'(1-y) + \frac{\tau}{2m_b^3} \delta'(1-y) \bigg]\,,
\end{eqnarray}
where 
\begin{equation}
y \equiv   \frac{2E_\ell}{m_b}.
\end{equation}
Here $3 \tau = -2({\cal T}_1 + 3 {\cal T}_3)$ and the $\rho_i$, ${\cal T}_i$ are the papameters used in \cite{GK,bauer}. This agrees with the corresponding expressions from Refs.~\cite{MW,GK}. 

We can now use these results to get the ${O}(\lqcd/m_B)$ corrections to the determination of $|V_{ub}|$ from the combined measurement of the semileptonic and radiative $B$ decays given in Eq.~(\ref{treeVub}). Using the differential decay rates given in Eqs.~(\ref{subleadingbsg}) and (\ref{finalbtou}), together with Eq.~(\ref{GuGsdefs}), we find
\begin{eqnarray}
\Gamma_u(E_c) &=& \frac{2\, \Gamma_0}{m_b}\int_{E_c}^{m_B/2}\!\!dE  \int_{-\infty}^{m_b-2E} \!\!d \omega \left[ F(\omega) 
\left(1 - \frac{\omega}{m_b}\right) 
\right.\nn\\
&&\left.
- \frac{1}{m_b} h_1(\omega) 
+ \frac{3}{m_b} H_2(\omega) \right]\\
\Gamma_s(E_c) &=& \frac{\Gamma_0^s}{m_b}\int_{E_c}^{m_B/2}\!\!dE  \int_{-\infty}^{m_b-2E} \!\!d \omega \left[ F(\omega) 
\left(1 - \frac{2\omega}{m_b}\right) 
\right.\nn\\
&&\left.
+ \frac{1}{m_b} h_1(\omega) 
+ \frac{1}{m_b} H_2(\omega) \right]
\end{eqnarray}
where $\Gamma_0^s$ and $\Gamma_0$ are defined in Eqs.~(\ref{Gamma0s}) and (\ref{Gamma0u}), respectively. 
This finally leads to 
\begin{widetext}
\begin{eqnarray}\label{Vubrelation}
\left|\frac{V_{ub}}{V_{tb}V_{ts}^*}\right| &=& 
\left(\frac{3 \alpha}{\pi} |C_7^{\rm eff}|^2 \frac{\Gamma_u(E_c)}{\Gamma_s(E_c)} 
\left[1 + 
\frac{1}{m_b} \frac{\int_{E_c}^{m_B/2}\!\!dE  \int_{-\infty}^{m_b-2E} \!\!d \omega  \big[ 2 h_1(\omega) - 2 H_2(\omega) - \omega f(\omega)\big]}
{\int_{E_c}^{m_B/2}\!\!dE  \int_{-\infty}^{m_b-2E} \!\!d \omega F(\omega)} \right]\right)^\frac12\nonumber\\
& \equiv&\left( \frac{3 \alpha}{\pi} |C_7^{\rm eff}|^2 \frac{\Gamma_u(E_c)}{\Gamma_s(E_c)}\right)^\frac12
\left(1 + \delta(E_c)\right),
\end{eqnarray}
\end{widetext}
where $\delta(E_c)$ is the correction to the extraction of $|V_{ub}|$ from subleading twist. 

We note from Eq.~(\ref{nonlocalexpand}) that the first moments of $h_1(\omega)$ and $H_2(\omega)$ have opposite signs, and so we expect their effects to add, not cancel, in the expression (\ref{Vubrelation}).  This gives the na\"\i ve estimate $\delta\sim 2\lqcd/m_b\sim 0.2$,  so the contributions of the subleading twist shape functions are sizable. 

\begin{figure}
\includegraphics[width=7cm]{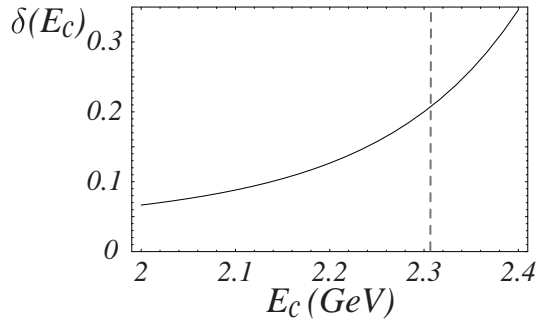}
\caption{The subleading twist corrections to the $|V_{ub}|$ relation (\ref{Vubrelation})  as a function on the lepton energy cut, using the simple model described in the text. The vertical line corresponds to the kinematical endpoint of the semileptonic $b \to c$ decay.\label{fig2}}
\end{figure}
To get a more quantitative estimate for these corrections, we can use the simple model for the subleading structure functions introduced in \cite{subleading}, in which these functions are taken as the derivative of the leading order shape function, normalized such that they reproduce the known first moments.  Using the one parameter model for the leading order structure function given in \cite{fmodel}
\begin{equation}
f_\model(\omega) = \frac{32}{\pi^2\Lambda} \left(1+\frac{\omega}{\Lambda}\right)^2
            \exp \left[ -\frac{4}{\pi} \left(1+\frac{\omega}{\Lambda}\right)^2 \right]\theta\left(1+\frac{\omega}{\Lambda}\right)
\end{equation}
(where $\Lambda=m_B-m_b$)
this gives
\begin{eqnarray}\label{Ffunct}
F_\model(\omega) &=& f_\model(\omega) + \frac{\lambda_1 -9 \lambda_2}{6\,m_b}
            f_\model'(\omega)  \nonumber\\
h_{1\,\model} &=& \lambda_2\, f_\model'(\omega)  \nonumber\\
H_{2\,\model} &=& - \lambda_2\, f_\model'(\omega)  
\end{eqnarray}
and leads to the expression for the subleading twist contributions
\begin{widetext}
\begin{eqnarray}
\delta_\model(E_c) = \frac{1}{2m_b} \, \frac{\int_{E_c}^{m_B/2}\!\!dE  \,\,\left[4\lambda_2f_\model(m_b-2E) - \int_{-\infty}^{m_b-2E} \!\!d \omega \,\, \omega f_\model(\omega)\right]}
{\int_{E_c}^{m_B/2}\!\!dE  \int_{-\infty}^{m_b-2E} \!\!d \omega \,\,f_\model(\omega)} \,.
\end{eqnarray}
\end{widetext}
Using the numerical value $\Lambda = 0.47\,\gev$ for the model parameter, we can estimate the subleading twist contributions $\delta$ as a function of the lepton energy cut $E_c$. This is shown in Fig.~\ref{fig2}.
%
%
For the cut used by the CLEO collaboration \cite{CLEObtou} to reject background from semileptonic $b \to c$ transitions $E_c = 2.2$ GeV, the subleading twist corrections to $|V_{ub}|$ are about 15\%. Since this number is strongly model dependent, we estimate the uncertainties in $|V_{ub}|$ to also be at the $\sim$ 15\% level. From Fig.~\ref{fig2} one can see that these uncertainties grow rapidly if the cut is raised further. Conversely, lowering the cut below 2.2 GeV reduces the uncertainty from
subleading twist contributions.  However, as the cut is lowered below 2.2
GeV the effect of the resummation of local operators in the twist expansion
becomes small and the usual local OPE in terms of local operators becomes
valid, eliminating the need to extract $f(w)$ from the $B \to X_s \gamma$ spectrum.

We would like to thank the staff of the Aspen Center for Physics
for their hospitality while some of this work was completed. This work was
supported by the Department of Energy under grants DOE-FG03-97ER40546 and
DOE-ER-40682-143 and the Natural Sciences and Engineering Research Council of
Canada, and by the DFG Graduiertenkolleg ``Hochenergiephysik and 
Teilchenastrophysik'', from the DFG Forschergruppe ``Quantenfeldtheorie,
Computeralgebra und Monte Carlo Simulationen'' and from the Ministerium f\"ur
Bildung und Forschung bmb+f.

\bigskip

\noindent {\it Note:  While this work was being completed, we became aware of similar results by Leibovich, Ligeti and Wise \cite{competition}. We are grateful to these authors for sharing their results with us before publication.}

\end{document}